# Electronic heat transport in the optimally doped superconductor $YBa_2Cu_3O_{6.95}$: preservation of the Wiedemann-Franz law

M. Matusiak [1,1], K. Rogacki [1], T. Plackowski [1], and B. Veal [2]

[1] *Institute of Low Temperature and Structure Research, Polish Academy of Sciences,*

*P.O. Box 1410, 50-950 Wrocław, Poland*

[2] *Materials Science Division, Argonne National Laboratory, Argonne, Illinois 60439, USA*

Measurements of longitudinal and transverse transport coefficients were performed in the normal state of a high-quality single crystal of optimally-doped superconductor $YBa_2Cu_3O_{6.95}$. We found the Hall-Lorenz number ($L_{xy}$) depends only weakly on temperature, thus no appreciable violation of the Wiedemann-Franz law has been observed. However, $L_{xy}$ is about two times larger than the Sommerfeld's value of the Lorenz number. Our results can be interpreted consistently if we assume that the electronic system of $YBa_2Cu_3O_{6.95}$ behaves like a Fermi-liquid with a pseudo-gap that opens at the Fermi level.

---

[1] Corresponding author. Tel.: +48-71-3435021; Fax: +48-71-3441029; e-mail: M.Matusiak@int.pan.wroc.pl

Knowledge of the behavior of electrons in the normal state of high temperature superconductors (HTSC) seems to be essential for understanding their unusual charge and heat transport properties as well as pairing mechanism in the superconducting state. In metals, charge carriers are well described in the framework of Fermi-liquid theory, which considers well defined fermionic "quasiparticles" with charge $e$, spin $1/2$, and an effective mass $m^*$ to be interaction dependent. One of the basic features of the Fermi liquid is a relation between thermal and electrical conductivities known as the Wiedemann-Franz (WF) law. At given temperature the WF law can be written in the convenient dimensionless form as:

$$L \equiv \frac{\kappa_{el}}{\sigma T}\left(\frac{e}{k_B}\right)^2, \qquad (1)$$

where $\kappa_{el}$ is the electronic thermal conductivity, $\sigma$ is the electrical conductivity, $k_B$ is Boltzmann's constant, and $L$ is the Lorenz number that in standard transport theory is equal to the Sommerfeld's value $L_0 = \pi^2/3$ if the mean-free-paths for transport of charge and entropy are identical. Deviations from this law can indicate that the Fermi liquid is not a ground state of the electronic system. Violation of the WF law has been recently reported in a cooper-oxide superconductor, $(Pr,Ce)_2CuO_4$, revealing that the elementary excitations that carry heat in this material at low temperatures are not fermions[1].

In general, the electronic thermal conductivity seems to be a quantity which can provide important information on the basic properties of the charge carriers and the entropy carriers as well. In $YBa_2Cu_3O_{7-d}$ (YBCO), advanced studies of the thermal conductivity have shown that a triangular pseudo-gap in the electron density of states may open at the Fermi level[2] and that an unconventional pairing symmetry of quasiparticles is possible in this compound[3,4]. Moreover, publications considering the nature of the ground state of cuprates[1,5] and the breaking of time reversal symmetry in $Bi_2Sr_2Ca(Cu_{1-x}Ni_x)_2O_8$[6] was also announced. However, the studies of the thermal conductivity in HTSC do not seem to be a trivial problem. The main difficulties refer to the interpretation of the experimental results, since heat in solids is usually carried by both electrons and phonons. The measured total thermal conductivity ($\kappa$) is the sum of the electronic ($\kappa_{el}$) and phonon ($\kappa_{ph}$) components, $\kappa = \kappa_{el} + \kappa_{ph}$, and a contribution from each component has to be determined. In conventional metals electrons carry most of heat, so the phonon contribution to the thermal conductivity can be in many cases neglected. However, in high-$T_c$ cuprates $\kappa_{ph}$ can be the same order of magnitude as $\kappa_{el}$ (e.g. in single crystals of YBCO[7] and $Bi_2Sr_2YCu_2O_8$[8]), or even 10 - 100 times larger (e.g. in polycrystalline samples of $REBa_2Cu_3O_7$[9]). Recently, Zhang et al.[10] used the thermal Hall effect to measure the electronic component separated from the total thermal conductivity. In zero magnetic field, the observed total thermal current is parallel to the applied temperature gradient $\nabla T$. In an external field perpendicular to $\nabla T$, a Lorentz force is generated that acts only on the electronic component of the heat current, because the phonon component is unaffected by the field. Thus, the transverse $\nabla T$ refers to the heat transport by the charge carriers only.

In this letter we study the thermal Hall effect in the optimally doped YBCO superconductor and present results on the determination of the Hall-Lorenz number ($L_{xy}$), which is regarded as a direct source of information about the electronic heat current[10]. Our studies have been performed in the normal-state of an $YBa_2Cu_3O_{6.95}$ single crystal and they show that $L_{xy}$ only weakly depends on temperature. In other words, we confirm validity of the WF law in the normal state of the optimally doped high temperature superconductor. The value of the Hall-Lorenz number, which significantly exceeds the Sommerfeld value at room temperature, is discussed and the presence of a pseudo-gap in the electronic structure is considered as an explanation.

The single crystals of $YBa_2Cu_3O_{7-d}$ were grown in a gold crucible by a conventional self-flux growth method. These were annealed in flowing oxygen for 72 hours while cooling from 500 C to 450 C. The magnetically measured superconducting transition of these crystals had an onset of $\approx 92.5 - 93.0$ K and a width of less than $\approx 1$ K as measured with 1 Oe ac-field amplitude perpendicular to the ab-plane. The crystal selected for measurements was about 1.55 mm long, 1.35 mm wide and 0.35 mm thick (along the crystallographic *c*-axis). This crystal was aged in air for a few years. The oxygen content was estimated to be equal to 6.95 (on the basis of the in-plane thermoelectric power value at room temperature $S_{ab}(300 K) \approx -1$ μV/K[11,12,13]). The resistively measured superconducting transition temperature was $T_c = 92.6$ K and the transition width was $\Delta T_c = 0.3$ K. The crystal was twinned, therefore the coefficients measured along the *ab*-plane should represent values averaged between the *a* and *b* crystallographic directions.

Electrical resistivity was measured using a four-point technique. The Hall coefficient measurements were performed by a standard method in a magnetic field of 13 T. The current and magnetic field directions were reversed several times to exclude any influence of the asymmetric position of the Hall contacts and detrimental electro-motive forces. The longitudinal and transversal thermal conductivities were measured in a single experiment. One edge of the crystal was glued to a heat sink, and a carbon heater was painted on the opposite edge. Using this heater a longitudinal temperature gradient ($\nabla_x T$) was generated. Due to the magnetic field applied perpendicularly to $\nabla_x T$ (and parallel to the *c* axis of the YBCO crystal) the transverse temperature gradient ($\nabla_y T$) appeared. Both gradients were measured with a pair of Chromel-Constantan thermocouples. Typically $\nabla_x T$ and $\nabla_y T$ were of the order of magnitude of 1 K/mm and 10 mK/mm, respectively. $\nabla_x T$ at $B = 0$ T was used to calculate the longitudinal thermal conductivity. We usually observed $\nabla_y T \neq 0$ at $B = 0$ T due to slightly asymmetric positions of the thermocouple junctions. This inconvenience was eliminated by varying the magnetic field between -13 and +13 T and by using, for calculations of the transversal thermal conductivity, only the slope $\Delta(\nabla_y T(B))/\Delta B$. Typically, $\Delta(\nabla_y T(B))/\Delta B$ was of the order of 0.1 mK/(mm T). To obtain credible results, $B$ was inverted several times with various $\nabla_x T$ at every considered temperature.

Figure 1 shows the temperature dependencies of four transport coefficients for the *ab*-plane of the $YBa_2Cu_3O_{6.95}$ single crystal. The longitudinal thermal and electrical conductivities ($\kappa_{xx}$ and $\sigma_{xx}$) as well as the transverse thermal and electrical conductivities ($\kappa_{xy}$ and $\sigma_{xy}$) were determined for the crystal aged for a few years. The measurements were performed within two weeks, thus the properties of the sample remained unchanged due to any possible long-term relaxation processes. Figure 1(a) shows the electrical resistivity ($\rho = 1/\sigma_{xx}$) and the longitudinal heat conductivity in the absence of a magnetic field. The striking feature of the $\rho(T)$ dependence is its perfect linearity between 100 K and 300 K, above the sharp transition to the superconducting state at $T_c = 92.6$ K. The value of $\rho$ at 300 K ($\approx 7$ μΩ m) is higher than that previously reported for optimally-doped YBCO single crystals, but may result from long term aging[14]. The thermal conductivity at room temperature is roughly the same as reported by other groups[15,16]. The value of $\kappa_{xx}$ slowly grows when the temperature falls to $T_c$.

The temperature dependencies of the Hall coefficient, $R_H = \sigma_{xy}/(\sigma_{xx})^2 B$, and transverse thermal conductivity, $\kappa_{xy} = \nabla_y T \cdot \kappa_{xx}/\nabla_x T$, are shown in Fig. 1(b) for the same crystal. The value of $R_H$ at 300 K is about two times larger than reported previously for optimally doped YBCO single crystals[17,18]. This is probably also caused by the long-time aging of our crystal, as mentioned above. On the other hand, the absolute value of $\kappa_{xy}$ at 300 K and its temperature dependence, $\kappa_{xy}(T)$, are almost identical to those obtained by Zhang et al. for untwined $YBa_2Cu_3O_{6.95}$[10]. We fitted our experimental data to the power dependence $\kappa_{xy}/B \sim T^\alpha$ with $\alpha = -1.21$, while Zhang et al. obtained $\alpha = -1.19$. These similarities in the absolute value and the temperature dependence of $\kappa_{xy}/B$, observed for twinned (our) and untwined crystals, reveal a rather low anisotropy of the thermal conductivity in the ab-plane. Striking and also puzzling is the similarity between $R_H(T)$ and $\kappa_{xy}(T)$.

The Hall-Lorenz number that is related only to the electronic thermal and charge conductivities can be calculated from the modified WF law[10]:

$$L_{xy} \equiv \frac{\kappa_{xy}}{\sigma_{xy}T}\left(\frac{e}{k_B}\right)^2. \qquad (2)$$

Within the Fermi-liquid approach the regular Lorenz number ($L$) as well as $L_{xy}$ are temperature independent and equal to the Sommerfeld's value if charge carriers scatter elastically[19]. There is an acceptable approximation at low temperatures where elastic scattering of electrons by impurities is dominant, and at high temperatures where electron-phonon scattering may be also regarded as elastic. In the intermediate temperature region, i.e. in the region where our studies were performed, the inelastic electron-phonon scattering more effectively disturbs the heat current than the charge current[20], and this results in a decrease of both Lorenz numbers[10]. In such a case, the different mean free paths for the transport of entropy ($l_s$) and charge ($l_e$) may be defined and $L_{xy}$ to $L$ compared by the ratio:

$$a_L \equiv \frac{L^2}{L_{xy}}, \qquad (3)$$

where $a_L$ is expected to be nearly constant, since $L \sim \langle l_s \rangle / \langle l_e \rangle$ and $L_{xy} \sim \langle l_s^2 \rangle / \langle l_e^2 \rangle$ [10,19]. When inelastic scattering appears, the Hall-Lorenz number falls faster than the regular Lorenz number, but both of these parameters stay correlated. This was found to be valid in copper[10]. In YBa$_2$Cu$_3$O$_{6.95}$ single crystals a dependence $L_{xy} \sim T$ was observed[10] and was interpreted as a violation of the WF law due to electron-electron scattering in the YBCO type high-$T_c$ superconductors. Our results contradict the strong $L_{xy}(T)$ dependence and hence also the violation of the WF law, as we discuss below.

Figure 2 shows the temperature dependence of the Hall-Lorenz number for our YBa$_2$Cu$_3$O$_{6.95}$ crystal in the normal state. Here, when temperature decreases, a rather small decline of $L_{xy}(T)$ is observed at intermediate temperatures but otherwise, remains unchanged or weakly temperature dependent. This is in excellent agreement with theoretical results obtained by Li in the framework of the marginal Fermi-liquid approximation[21]. The weak $L_{xy}(T)$ dependence means that charge carriers in YBCO approximately follow the WF law. Our $L_{xy}(T)$ results obtained for optimally doped YBa$_2$Cu$_3$O$_{6.95}$ differ significantly from those reported in Ref. 10. Any reasonable extrapolation of the $L_{xy}(T)$ data in Fig. 2, even for a linear extrapolation, gives $L_{xy} \geq 3$ at $T = 0$, i.e. a value much higher than $L_{xy} \approx 0$ obtained in Ref. 10. A possible reason (suggested in ref. 21) for the significant differences in the $L_{xy}(T)$ dependencies reported here and in ref. 10 might be that different amounts of out-of-plane impurities existed in the two YBa$_2$Cu$_3$O$_{6.95}$ crystals used in Ref. 10, one for $\kappa_{xy}$ and the second for $\sigma_{xy}$ measurements. This has been discussed in detail in Ref. 21. To avoid potential serious inconsistencies, results presented in this work have been obtained from measurements performed on the same crystal.

Similar to our YBCO results, a weak $L_{xy}(T)$ dependence has also been observed for an optimally doped non-aged EuBa$_2$Cu$_3$O$_7$ (Eu123) single crystal[22]. The data obtained for Eu123 are shown in Fig.2 to compare with our results for YBCO. Similar results were also observed for an underdoped Eu123 crystal in a temperature range 160 - 300 K (and will be published separately[22]). We believe that the strong similarity in $L_{xy}(T)$ for these different systems suggests universal behavior for the temperature dependence of $L_{xy}$ in YBCO-type superconductors.

The temperature dependence of the electronic contribution to the total thermal conductivity may be derived if we assume that at room temperature the scattering of electrons is approximately elastic. Then, $\kappa_{el}(T)$ can be obtained by using an assumption $L(300) \approx L_{xy}(300)$, Eq. 3 and the WF law (Eq. 1). The results are shown in the inset of Fig. 2. With decreasing temperature, $\kappa_{el}(T)$ goes down and this may be interpreted as a result of the decreased concentration of the charge carriers. Similarities between the behavior of $\kappa_{el}(T)$ and the temperature dependence of the Hall-concentration ($n_H = 1/(e R_H)$), that is shown in the same inset, support this conclusion.

Results obtained for transverse thermal and electrical conductivities can be well fitted with the functions derived in the framework of the marginal-Fermi-liquid hypothesis[21]: $\kappa_{xy}(T) = a_h T^{-1} + b_h T^{-2}$ and $\sigma_{xy}(T) = a_c T^{-2} + b_c T^{-3}$. However, the fitting parameters remain in the ratio $b_h/a_h \approx 0.13\, b_c/a_c$, which is much lower than $b_h/a_h \approx 1.4\, b_c/a_c$, obtained in Ref. 21. This large discrepancy between experimental and theoretical results may have some meaning for farther modification or development of the marginal-Fermi-liquid hypothesis. Another effort to describe electrical and thermal transport in the normal state of HTSC was made by Coleman et al.[23], who based their calculations on Anderson's model with two different transport relaxation times of quasiparticles[24]. They reconsidered Anderson's model and proposed an original modification of the normal Fermi-liquid behavior. They argued that in cuprates the thermal transport can be understood in terms of scattering processes that are sensitive to the charge-conjugation symmetry of the quasiparticles. Coleman et al. obtained $\kappa_{xy} \sim T^{-2}$ and $\sigma_{xy} \sim T^{-3}$; i.e., dependencies that are different from our experimental results: $\kappa_{xy} \sim T^{-1.2}$ and $\sigma_{xy} \sim T^{-2.7}$.

The above mentioned discrepancies between our experiments and predictions of the marginal- and non- Fermi-liquid theories[21,23] show that no convincing evidence has been found in these experiments that the charge carriers in the normal state of $YBa_2Cu_3O_{6.95}$ behave much differently than the regular Fermi-liquid. The remaining question is why the value of $L_{xy}$ is higher than the Sommerfeld value of the Lorenz number ($L_0 \approx 3.3$). Despite the systematic error that can be as large as 30% of the obtained value (mainly due to sample geometry uncertainties), the smallest possible $L_{xy}$ at high temperatures is definitely larger than $L_0$. This can not be explained within the standard Fermi-liquid theory. However, an increase of $L$ is expected if the pseudo-gap that opens at the Fermi surface is taken into account, as has been recently shown by Minami et al.[2]. They considered the triangular pseudo-gap and found that the ratio $L / L_0$ can reach a value of $\approx 2.5$, depending on temperature and width of the pseudo-gap. This agrees with our data obtained above $T_c$. Additionally, our results are consistent with experimental studies of the thermal transport in other HTSC at temperatures lower than $T_c$ and in high magnetic fields, that also suggest that the Lorenz number can significantly exceed the Sommerfeld's value[1,25].

In summary, we have investigated transport properties in the normal state of a superconducting $YBa_2Cu_3O_{6.95}$ single crystal. We have studied the temperature dependencies of the longitudinal electrical and thermal conductivities, as well as the Hall and thermal Hall effects represented by the transverse electrical and thermal conductivities, respectively. The obtained results have been used to calculate the temperature dependence of the Hall-Lorenz number ($L_{xy}$). This number varies slowly with temperature and is about two times larger than the Sommerfeld value of the Lorenz number. We have concluded that the normal state of $YBa_2Cu_3O_{6.95}$ can be understood within the Fermi-liquid approximation when modified by the presence of a pseudo-gap at the Fermi level.

**Acknowledgements**

The authors are grateful to Dr. C. Sułkowski for thermopower measurements. Support by U.S. Dept. of Energy under contract #W-31-109-ENG-38 (BWV) and by the Polish State Committee for Scientific Research under contract No. 2 POB 036 24 (TP) is acknowledged.

**Figure captions**

1. The temperature dependences of the transport coefficients for the $YBa_2Cu_3O_{6.95}$ single crystal. In the upper panel (*a*) the longitudinal electrical resistivity (solid line and left axis) and the longitudinal thermal conductivity (open points and right axis) are presented; the dashed line shows the linear fit of the resistivity in a range from 100 to 300 K. The bottom panel (*b*) shows the Hall coefficient (solid line and left axis) and the transverse thermal conductivity divided by the magnetic field (open points and right axis); the dashed line shows a function $aT^{-1.21}$ being the best power fit to the experimental points.

2. The temperature dependence of the Hall-Lorenz number for the $YBa_2Cu_3O_{6.95}$ single crystal; the dashed line is a guide for the eye. The x's present values of $L_{xy}$ obtained for an optimally doped $EuBa_2Cu_3O_{7-\delta}$ single crystal[22]. The inset shows the temperature dependences of the Hall-concentration (solid line and left axis) and the electronic thermal conductivity (open points and right axis) for the $YBa_2Cu_3O_{6.95}$ crystal.

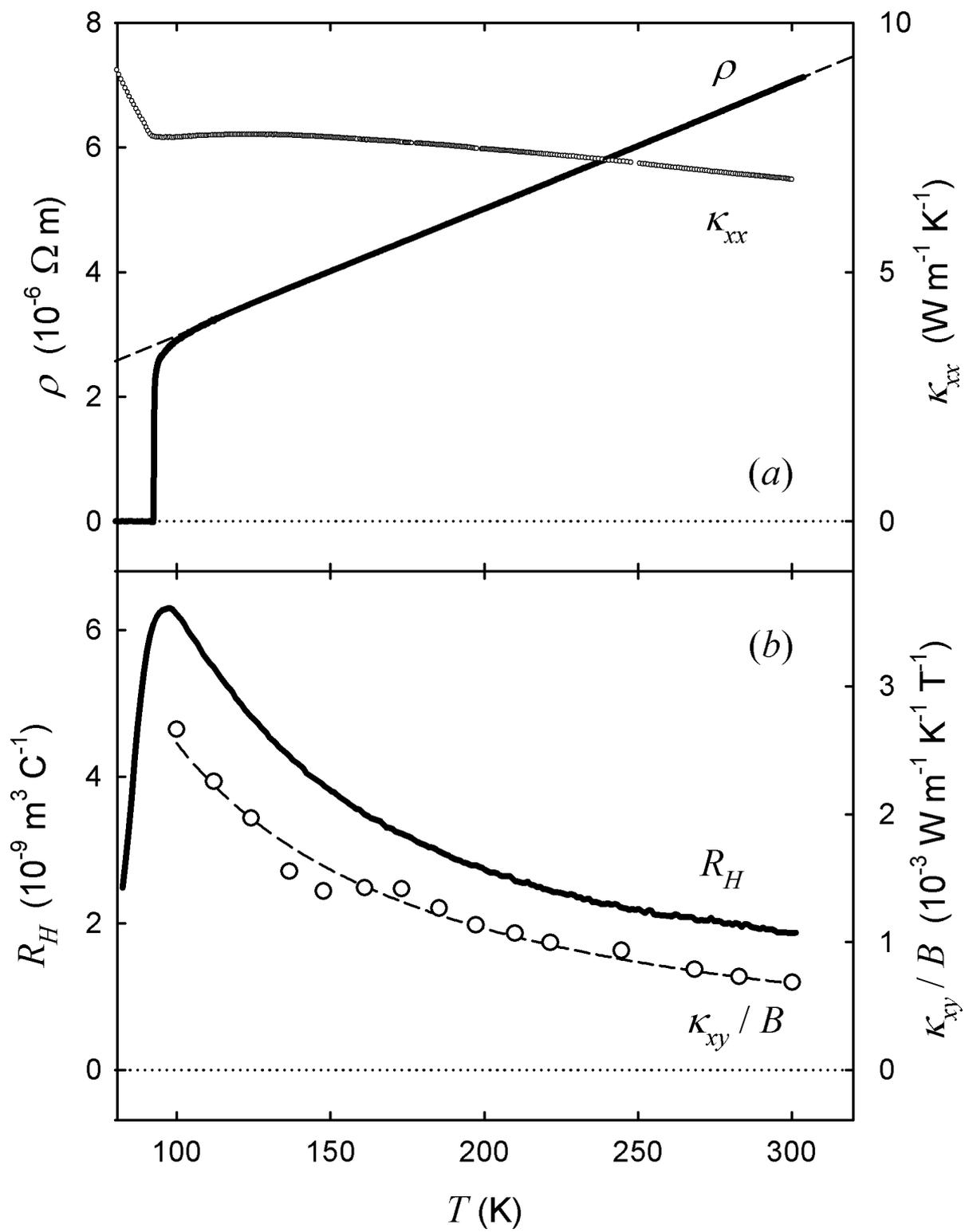

Figure 1.

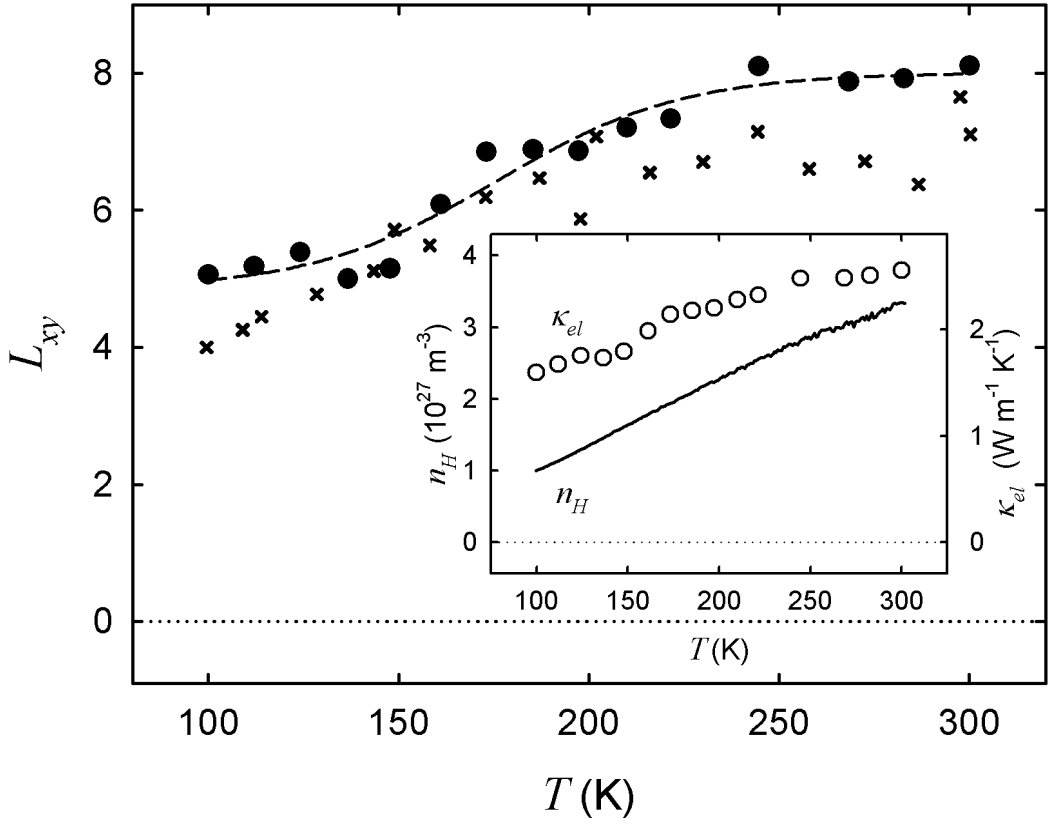

**Figure 2.**